\documentclass{revtex4}
\usepackage{epsfig}
\usepackage{amssymb}
\usepackage{amsmath}
\usepackage{amsfonts}
\usepackage{graphicx}
\usepackage{mathrsfs}
\usepackage[dvips]{color}
\usepackage{multirow}

\newcommand{\fa}{\mathfrak{a}}

\newcommand{\fu}{\mathfrak{u}}

\newcommand{\fn}{{\,\mathfrak{n}\,}}

\newcommand{\fz}{\mathfrak{z}}

\newcommand{\fK}{\mathfrak{K}}

\newcommand{\bM}{\mathbf{M}}

\newcommand{\cB}{\mathcal{B}}

\newcommand{\cD}{\mathcal{D}}
\newcommand{\cE}{\mathcal{E}}
\newcommand{\cH}{\mathcal{H}}

\newcommand{\cJ}{\mathcal{J}}

\newcommand{\cP}{\mathcal{P}}

\newcommand{\cT}{\mathcal{T}}

\newcommand{\be}{\begin{equation}}
\newcommand{\ee}{\end{equation}}
\newcommand{\bea}{\begin{eqnarray}}
\newcommand{\eea}{\end{eqnarray}}
\newcommand{\nn}{\nonumber}

\newcommand{\ed}{\end{document}}

\newcommand{\bi}{\begin{itemize}}
\newcommand{\ei}{\end{itemize}}

\newcommand{\bce}{\begin{center}}
\newcommand{\ece}{\end{center}}

\newcommand{\sE}{\mathscr{E}}

\begin{document}

\title{$\mathcal{P}\mathcal{T}$-Symmetric Coherent Perfect Absorber with Graphene}

\author{Mustafa Sar{\i}saman$^1$}\email{mustafa.sarisaman@istanbul.edu.tr}
\affiliation{Department of Physics, Istanbul University, 34134 Istanbul, Turkey}
\author{Murat Tas$^2$}\email{tasm236@gmail.com}
\affiliation{Department of Physics, Gebze Technical University, 41400 Kocaeli, Turkey}

\begin{abstract}

We investigate $\mathcal{PT}$-symmetric coherent perfect absorbers (CPAs) in the TE mode solution of a linear homogeneous optical system surrounded by graphene sheets. It is revealed that presence of graphene sheets contributes the enhancement of absorption in a coherent perfect absorber. We derive exact analytic expressions, and work through their possible impacts on lasing threshold and CPA conditions. We point out roles of each parameter governing optical system with graphene and show that optimal conditions of these parameters give rise to enhancement and possible experimental realization of a CPA laser. Presence of graphene leads the required gain amount to reduce considerably based on its chemical potential and temperature. We obtain that relation between system parameters decides the measure of CPA condition. We find out that graphene features contributing to resonance effect in graphene sheets are rather preferable to build a better coherent perfect absorber.
\medskip

\noindent {Pacs numbers: 03.65.Nk,  42.25.Bs, 42.60.Da,
24.30.Gd}
\end{abstract}

\maketitle

\section{Introduction}

The coherent perfect absorbers (CPAs), or antilasers,~\cite{antilaser1,antilaser2,antilaser2-1,antilaser2-2,
antilaser2-3,antilaser3,antilaser4,antilaser5} are fascinating optical constructs that furnish in principle
the absorption of certain incident coherent waves by the optical potential, but yet they basically act as a
theoretical design today. Since they operate as time reversal case of regular lasers, they are inherently
expressed by the time reversal symmetry of spectral singularities~\cite{naimark,naimark-1,naimark-2,
naimark-3,naimark-4}, which are used as a tool to render the best way of examining lasing threshold
condition~\cite{silfvast} in a laser. Thus, complex conjugate of the optical potential that supports a
spectral singularity engenders a CPA-laser action~\cite{jpa-2012}. Accordingly, it is intriguing to consider
a $\cP\cT$-symmetric potential that endorses both a spectral singularity and CPA concurrently. This is rather
interesting because it functions as a laser emitting coherent waves unless it is subject to incident coherent
waves with appropriate amplitude and phase in which case it acts as an absorber~\cite{lastpaper}. Although
recent experimental realizations of CPAs have been carried out~\cite{CPAexp,CPAexp-1,CPAexp-2}, basically
they still stand as theoretical structures awaiting experimental advancement, see also~\cite{antilaser5} for
a recent review of CPAs.

Since its debut, $\cP\cT$ symmetry has found considerable interest in optics and related fields due to its
smoothness to realize experimental investigations and immediate applications. A generic $\cP\cT$-symmetric
Hamiltonian in optics is vested with a potential whose peculiar property is $V(x)= V^{\star}(-x)$~\cite{
bender,bender-1,bender-2,bender-3,bender-4,PT4,PT5}, which corresponds to switching gain and loss components
in conserved potential. Complex optical $\cP\cT$-symmetric potentials are realized by the formal equivalence
between quantum mechanical Schr\"{o}dinger equation and optical wave equation derived from Maxwell equations.
By exploiting optical modulation of the refractive index in the complex dielectric permittivity plane and
engineering both optical absorption and amplification, $\cP\cT$-symmetric optical systems can lead to a
series of intriguing optical phenomena and devices, such as dynamic power oscillations of light propagation,
CPA-lasers~\cite{lastpaper,CPA,CPA-1,CPA-2,CPA-3,pra-2015d}, spectral singularities~\cite{naimark,naimark-1,
naimark-2,naimark-3,naimark-4,p123,pra-2012a,longhi4,longhi3} and unidirectional invisibility~\cite{bender,
bender-1,bender-2,bender-3,bender-4,PT6,PT7,pra-2017a,prb-2018}.

Emergence of CPA-lasers is among the most notable applications of $\cP\cT$-symmetric potentials in optics.
In the context of $\cP\cT$-symmetry, the condition for appearance of a spectral singularity coincides with
that of its time-reversal~\cite{longhi2010}. This makes $\cP\cT$-symmetric CPA-lasers as one of the primary
examples in the study of $\mathcal{PT}$-symmetric optical structures~\cite{mostafazadeh2012}. This elegant
finding motivates new insight towards solving the problem of constructing a CPA with appropriate amplitudes
and phases of incoming waves. In view of this motivation, here we examine feasibility of realizing a
CPA-laser in a homogeneous $\mathcal{PT}$-symmetric optical slab system covered by graphene sheets. Our aim
in using the graphene sheets is to enhance the adjustment of absorption in entire system~\cite{graphene,
graphene-1,graphene-2,graphene-3,graphene-4,graphene-5}.

Physical properties of graphene has been profoundly unveiled and thus its numerous applications in condensed
matter physics and optics have attracted interest of researchers for over a decade~\cite{gr1,gr2,gr3,gr4}.
Since its early discovery, a voluminous literature has arisen and plenty of applications have been realized
especially in the fields of sensor based transport phenomena~\cite{gr5,gr5-1,gr5-2,gr5-3,gr5-4,gr5-5,gr5-6},
impurity invisibility~\cite{gr6}, electron optics with p-n junctions~\cite{gr7}, and invisibility
cloaking~\cite{gr8,gr9,naserpour}. The idea that graphene interact with electromagnetic waves in anomalous
and exotic ways, providing new phenomena and applications, gives rise to the study of CPA phenomenon in
$\cP\cT$-symmetric optical structures with graphene. Especially recent works in this field~\cite{lastpaper,
graphene,graphene-1,graphene-2,graphene-3,graphene-4,graphene-5} fashion up essential motivation for this
study, which will use whole competency of the transfer matrix method in a scattering formalism which grounds
its power on Maxwell equations. Furthermore, see~\cite{referee1,referee2,referee3,referee4} for some
relevant studies of graphene based light modulation and optical gain.

In a scattering problem the transfer matrix is used to encapsule all the scattering data~\cite{prl-2009}.
Its composition property makes it more practical and preferable than scattering matrix. Moreover, its
components involve necessary information to reveal spectral singularities, CPA and invisibility of
electromagnetic fields interacting with an optically active medium~\cite{lastpaper,CPA,CPA-1,CPA-2,CPA-3,
pra-2015d}. See also \cite{transmat1,transmat2,transmat3} for the use of transfer matrix formalism
belonging to the scattering of layered structures containing graphene.

\begin{figure}[!h]
    \begin{center}
    \includegraphics[scale=0.5]{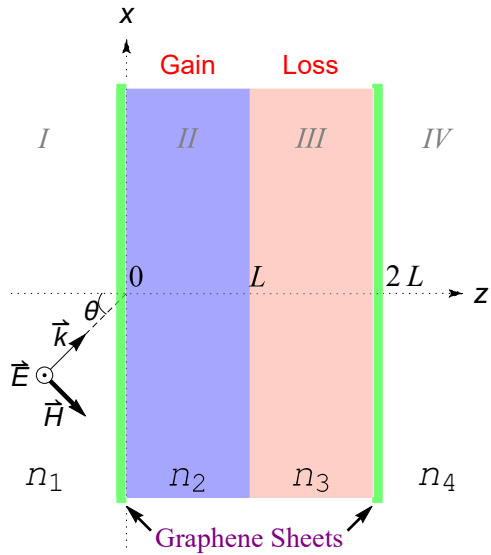}
    \caption{(Color online) Diagram depicting the TE mode solution of parallel pair of slab system covered
by graphene sheets in vacuum. Regions $I\!I$ and $I\!I\!I$ with thickness $L$ are respectively the gain and
loss layers having refractive indices $\fn_2$ and $\fn_3$. Regions $I$ and $I\!V$ correspond to the vacuum.}
    \label{fig1}
    \end{center}
    \end{figure}

Our analysis fulfils oblique incidence~\cite{lastpaper,jones1}, considering that desired phenomena may be
angle dependent. Hence, we conduct a comprehensive study of spectral singularities which yields lasing
threshold condition and CPAs in the oblique TE mode of a $\cP\cT$-symmetric system with graphene to unveil
the intriguing traits of transfer matrix as complementary to~\cite{lastpaper}. Our system is depicted in
Fig.~\ref{fig1}.

Our analysis demonstrates all possible configurations of the system leading to spectral singularity and CPA
solutions. We find out complete solutions and schematically demonstrate their behaviors using various
parameter choices. Among all possible parameters of the system which provide valuable information about the
CPA-lasers, only optimal ones lead to achieve performing an efficient CPA. In particular, we obtain analytic
expressions for the spectral singularity and CPA configurations, and examine behaviors of practically most
desirable choices of parameters corresponding to TE waves. We reveal that optimal control of graphene
parameters, such as gain coefficient, incidence angle, slab thickness, temperature, and chemical potential,
gives rise to a desired outcome for achieving enhancement of absorption, and computing correct amplitude
and phase contrasts in a CPA-laser. Thus, we provide a concrete ground that restrict the mentioned
parameters of graphene in certain ranges. Optimal values of these parameters should be adjusted in a given
system if one desires the experimental realization of CPA.

\section{TE Mode Solution of a Planar Bilayer Slab System with Graphenes}\label{S2}

Consider a one-dimensional linear homogeneous and optically active parallel pair of slab system whose
exterior planar surfaces are covered by graphene layers as depicted in Fig.~\ref{fig1}. Suppose that
entire optical system is immersed in air and regions $I\!I$ and $I\!I\!I$ are respectively filled with gain
and loss materials having constant complex refractive indices $\fn_2$ and $\fn_3$. Let this system be
exposed to external time harmonic electromagnetic waves with electric $\vec \cE$ and magnetic $\vec \cH$
fields. Maxwell equations describing interaction of the electromagnetic waves with the slab system have the
form:
    \begin{align}
    &\vec{\nabla}\cdot\vec{\cD}^{j} = \rho_j (z), &&
    \vec{\nabla}\cdot\vec{\cB}^{j} = 0,
    \label{equation1}\\
    &\vec{\nabla} \times \vec{\cH}^{j}-\partial_t \vec{\cD}^{j}=\sigma_j(z) \vec{\cE}^{j}, &&
    \vec{\nabla} \times \vec{\cE}^{j} +\partial_t \vec{\cB}^{j}= 0,
        \label{equation2}
    \end{align}
where the index $j=1,2,3,4$ represents the regions sketched in Fig.~\ref{fig1}, $\vec \cE^{j}$ and
$\vec \cH^{j}$ are the electric and magnetic fields in corresponding regions. They are connected to
$\vec \cD^{j}$ and $\vec \cB^{j}$ fields via the constitutive relations
    \begin{align*}
    \vec{\cD}^{j} := \varepsilon_0 \fz_{j}(z)\, \vec{\cE}^{j}, &&\vec{\cB}^{j}:=\mu_0\vec{\cH}^{j},
    \end{align*}
$\varepsilon_0$ and $\mu_0$ are respectively the permeability and permittivity of the vacuum. We defined
complex quantity $\fz_{j}(z)$
    \begin{align}
    \fz_{j}(z):= \fn_j ~~~~~ {\rm for~}z\in z_{j},
    \label{e1}
    \end{align}
such that the subindex $j=1,4$ correspond to vacuum at which $\fn_1=\fn_4=1$, $j=2,3$ represent respectively
the gain and loss components of the slab, and $z_{j}$ stands for $z$ coordinates in the specified $j$-th
region as depicted in Fig.~\ref{fig1}. In Maxwell equations (\ref{equation1}) and (\ref{equation2}),
$\rho_j(z)$ and $\sigma_j(z)$ respectively denote the free charge and conductivity present on the graphene
sheets and therefore expressed as
    \begin{align*}
    &\rho_j(z) := \rho_g^{(1)} \delta(z) + \rho_g^{(2)} \delta(z-2L),\\ &\sigma_j(z) := \sigma_g^{(1)} \delta(z) + \sigma_g^{(2)} \delta(z-2L),
    \end{align*}
where $\rho_g^{(\ell)}$ and $\sigma_g^{(\ell)}$ are respectively the free charge and conductivity on the
$\ell$-th layer of graphene, with $\ell=1,2$. Notice that $\rho_j(z)$ and $\sigma_j(z)$ are associated to
each other by the continuity equation
    \be
    \vec{\nabla}\cdot\vec{\cJ}_j + \partial_t \rho_j(z) = 0
    \ee
for the electric current density $\vec{\cJ}_j := \sigma_j(z) \vec{\cE}^j$. Conductivity of graphene sheets has
been determined within the random phase approximation in~\cite{conductivity-graphene,conductivity-graphene-1,
conductivity-graphene-2} as the sum of intraband and interband contributions, i.e.
$\sigma_{g} = \sigma_{intra}+\sigma_{inter}$, where
   \begin{align}
   \sigma_{intra} &:= \left[\left. ie^2\chi \right/ \pi\hbar^2\left(\omega + i\Gamma\right) \right]\ln\left[2\cosh\left(\frac{\mu}{\chi}\right)\right], \nn\\
   \sigma_{inter} &:=\frac{e^2}{4\pi\hbar}\Bigl[\frac{\pi}{2} + \arctan\left(\frac{\nu_{-}}{\chi}\right) -\frac{i}{2}\ln\frac{\nu_{+}^2}{\nu_{-}^2 + \chi^2}\Bigr].
   \label{conductivitydefns}
   \end{align}
Here $\nu_{\pm}:=\hbar\omega \pm 2\mu$, $\chi:=2k_B T$, $-e$ is electron charge, $\hbar$ is reduced Planck's
constant, $k_B$ is Boltzmann constant, $T$ is temperature, $\Gamma$ is the scattering rate of charge carriers,
$\mu$ is the chemical potential, and $\hbar\omega$ is the photon energy~\cite{naserpour}. In time harmonic
forms, $\vec \cE^{j}(\vec{r},t)$ and $\vec \cH^{j}(\vec{r},t)$ fields are respectively given by
$\vec \cE^{j}(\vec{r},t)=e^{-i\omega t} \vec{E}^{j}(\vec{r})$ and $\vec \cH^{j}(\vec{r},t)=e^{-i\omega t}
\vec{H}^{j}(\vec{r})$. Thus, Maxwell equations corresponding to transverse electric (TE) wave solutions yield
the following form of Helmholtz equation
    \begin{align}
    &\left[\nabla^{2} +k^2\fz_{j}(z)\right] \vec{E}^{j}(\vec{r}) = 0, &&
    \vec{H}^{j}(\vec{r}) = -\frac{i}{k Z_{0}} \vec{\nabla} \times \vec{E}^{j}(\vec{r}),
    \label{equation4}
    \end{align}
where $\vec r:=(x,y,z)$, $k:=\omega/c$ is the wavenumber, $c:=1/\sqrt{\mu_{0}\varepsilon_0}$ is the speed of
light in vacuum, and $Z_0:=\sqrt{\mu_0/\varepsilon_0}$ is the impedance of vacuum. We stress out that TE waves
correspond to the solutions of (\ref{equation4}) for which $\vec \cE^{j}(\vec{r})$ is parallel to the surface
of the slabs. In our geometrical setup, they are aligned along the $y$-axis. Suppose that in region $I$,
incident wave $\vec E^{1}(\vec{r})$ adapts a plane wave with wavevector $\vec k$ in the $x$-$z$ plane,
specified by
    \begin{align}
    &\vec k=k_x \hat e_x+ k_z \hat e_z, && k_x:=k\sin\theta, &&k_z:=k\cos\theta,
    \end{align}
where $\hat e_x,\hat e_y,$ and $\hat e_z,$ are respectively the unit vectors along the $x$-, $y$- and
$z$-directions, and $\theta\in[-90^\circ,90^\circ]$ is the incidence angle (See Fig.~\ref{fig1}). For
convenience we introduce the scaled variables
    \begin{align}
    &\mathbf{z}:=\frac{z}{L}, &&\mathbf{x}:=\frac{x}{L},  &&\fK:=Lk_z=kL\cos\theta . \label{scaled-var}
    \end{align}
Thus, the electric field corresponding to TE waves is given by
    \begin{align}
    \vec E^{j}(\vec{r})=\sE^{j} (L \mathbf{z})e^{i\fK \mathbf{x} \tan\theta}\hat e_y,
    \label{ez1}
    \end{align}
where $\sE^{j}$ is solution of the Schr\"odinger equation
    \be
    -\psi^{j''}(\mathbf{z})+v_{j}(\mathbf{z})\psi^{j}(\mathbf{z})=\fK^2\psi^{j}(\mathbf{z})~~~~~~~~~~\mathbf{z}\notin\{ 0,1,2\},
    \label{sch-eq}
    \ee
for the potential $v_{j}(\mathbf{z}):=\fK^2[1-\tilde\fn_j^2]$. Here we define $\tilde{\fn}_j$
    \be
    \tilde\fn_j:=\sec\theta \sqrt{\fn^2_j -\sin^2\theta}.\label{ntilde}
    \ee
The fact that potential $v_j(\mathbf{z})$ is constant in regions of interest gives rise to a solution in
relevant regions
    \be
    \psi^{j}(\mathbf{z}):=a_j\,e^{i{\fK}_j \mathbf{z}} + b_j\,e^{-i{\fK}_j \mathbf{z}} ~~~~~~ {\rm for} ~~ \mathbf{z}\in \mathbf{z}_{j},
    \label{E-theta}
    \ee
where $a_j$ and $b_j$, with $j=1,2,3,4$, are $\fK$-dependent complex coefficients, and
    \begin{align}
    &{\fK}_j:=\fK \tilde\fn_j .
    \label{tilde-parm}
    \end{align}
In particular, $\sE^j(L \mathbf{z})$ is given by the right-hand side of (\ref{E-theta}) with generally
different choices of constants $a_j$ and $b_j$. These coefficients are related to each other via appropriate
boundary conditions: tangential components of $\vec E^j$ and $\vec H^j$ are continuous across the surface
while the normal components of $\vec H^j$ have a step of unbounded surface currents across the interface of
graphenes. Table~\ref{table01} displays corresponding set of boundary conditions. Quantities $\fu_{\pm}^{(j)}$
are defined as
     \be
     \fu_{\pm}^{(j)} := \frac{1\pm \sigma_g^{(j)}}{\tilde\fn_j}.
     \label{u=}
     \ee

 \begin{table}[!htbp]
    \begin{center}
	{
    \begin{tabular}{|c|c|}
    \hline
    &\\[-10pt]
    $\mathbf{z}=0$ &
    $\begin{aligned}
    & a_2+ b_2 = a_1 + b_1 , && a_2 - b_2 = \fu_{+}^{(2)} a_1- \fu_{-}^{(2)} b_1\\[3pt]
    \end{aligned}$\\
    \hline
    &\\[-10pt]
    $\mathbf{z}=1$ & $\begin{aligned}
    & a_2 e^{i{\fK}_2} + b_2 e^{-i{\fK}_2}= a_3 e^{i{\fK}_3} + b_3 e^{-i{\fK}_3} \\[3pt]
    & \tilde\fn_2 (a_2 e^{i{\fK}_2} - b_2 e^{-i{\fK}_2})=
    \tilde\fn_3(a_3 e^{i{\fK}_3} - b_3 e^{-i{\fK}_3})
    \end{aligned}$\\[-8pt]
    &\\
    \hline
     &\\[-10pt]
     $\mathbf{z}=2$ & $\begin{aligned}
    & a_3 e^{2i{\fK}_3} + b_3 e^{-2i{\fK}_3}= a_4 e^{2i\fK} + b_4 e^{-2i\fK} \\[3pt]
    & a_3 e^{2i{\fK}_3} - b_3 e^{-2i{\fK}_3}=
   \fu_{-}^{(3)} a_4 e^{2i\fK} - \fu_{+}^{(3)} b_4 e^{-2i\fK}
    \end{aligned}$\\[-8pt]
    &\\
    \hline
    \end{tabular}}
    \vspace{6pt}
    \caption{Boundary conditions for TE waves.}
    \label{table01}
    \end{center}
\end{table}

\section{Transfer Matrix and Spectral Singularities}\label{S3}

Right outgoing waves could be associated to the left ones by means of the transfer matrix. Transfer matrix is
favored over the scattering matrix by virtue of its composition property, which helps being articulated the
scattering properties of any optical system. For our two-layer optical system furnished by $\cP\cT$-symmetry
with graphene sheets, total transfer matrix can be obtained as the product of transfer matrices of gain and
loss regions. If individual transfer matrices corresponding to the gain and loss regions of the slab are
denoted by $\bM_{1}$ and $\bM_2$ respectively, then total transfer matrix $\bM=[M_{ij}]$ satisfies
composition property $\bM=\bM_2\bM_1$, and is expressed by
     \begin{align}
    \left[\begin{array}{c}
    a_4\\ b_4\end{array}\right]=\bM \left[\begin{array}{c}
    a_1\\ b_1\end{array}\right].
    \nn
    \end{align}
Spectral singularities match up to real zeros of $M_{22}$ component of $\bM$, which is computed explicitly as
    \be
    M_{22} =\frac{e^{2i\fK}}{8}\left[V_{+}(\fu_{-}^{(3)}-1)e^{i{\fK}_3} + V_{-}(\fu_{-}^{(3)}+1)e^{-i{\fK}_3 }\right],
    \label{M22=x}
    \ee
where we identify
    \be
    V_{\pm} := (\tilde{\fn}_3 \pm \tilde{\fn}_2) (1 - \fu_{-}^{(2)})e^{i{\fK}_2} + (\tilde{\fn}_3 \mp \tilde{\fn}_2) (1 + \fu_{-}^{(2)})e^{-i{\fK}_2}.
    \notag
    \ee
It is apparent that $\cP\cT$-symmetry implies the following relations
    \begin{align}
    &\fn_2\stackrel{\mathcal{PT}}{\longleftrightarrow}\fn_3,
    &&\tilde{\fn}_2\stackrel{ \mathcal{PT} }{\longleftrightarrow}\tilde{\fn}_3,\nn\\
    &\fu_{\pm}^{(2)}\stackrel{ \mathcal{PT} }{\longleftrightarrow}\fu_{\mp}^{(3)},
    &&\sigma_g^{(2)}\stackrel{ \mathcal{PT} }{\longleftrightarrow}-\sigma_g^{(3)}.
    \label{pt-symmetry-rels}
    \end{align}
Thus it amounts that currents on the left and right graphene sheets flow in opposite directions by virtue of
$\cP\cT$ symmetry. Spectral singularities correspond to real values of the wavenumber $k$ such that $M_{22}=0$.
Hence, (\ref{M22=x}) gives rise to explicit form of the spectral singularity condition
    \be
    e^{2i{\fK}_3} = \frac{V_{-} \left(1 + \fu_{-}^{(3)}\right)}{V_{+} \left(1 - \fu_{-}^{(3)}\right)}.
    \label{spec-sing}
    \ee
We note that quantities $\fu_{-}^{(2,3)}$ involve the effect of graphene in the spectral singularities as
given by identity (\ref{u=}). Provided that graphene layers are removed by setting $\sigma_{g}^{(2,3)}=0$,
(\ref{spec-sing}) generates the spectral singularity condition given in~\cite{lastpaper}.

\section{Spectral Singularities in $\cP\cT$-Symmetric Configurations}\label{S4}

Spectral singularities corresponding to our optical setup in (\ref{spec-sing}) describe the lasing threshold
condition. This is in fact a complex expression screening behavior of system parameters. Thus, it can be
explored in detail by means of relevant quantities containing significant physical consequences. In view of
$\cP\cT$ symmetry relations given in (\ref{pt-symmetry-rels}), we obtain the following relations
\begin{align}
    &\fn:= \fn_2 =\fn_3^*,
    &&\tilde\fn:= \tilde\fn_2 =\tilde\fn_3^*, \notag\\
    &\sigma_g:= \sigma_g^{(2)}= -\sigma_g^{(3)\ast},
    &&\fu_{\pm}:=\fu_{\pm}^{(2)}= \fu_{\mp}^{(3)\ast}.
    \label{eq251}
    \end{align}
Refractive index $\fn$ and $\tilde{\fn}$ are expressed in real and imaginary parts as follows
    \begin{align}
    &\fn = \eta + i \kappa,
    &&\tilde\fn = \tilde\eta + i \tilde\kappa.
    \label{eq252}
    \end{align}
Most of the materials safely satisfy the condition $\left| \kappa \right| \ll \eta - 1 < \eta$. In particular,
this practical restriction of materials leads $\tilde\eta$ and $\tilde\kappa$ to be expressed in terms of
$\eta$ and leading order of $\kappa$ as
    \begin{align}
    &\tilde\eta \approx \sec\theta\sqrt{\eta^2-\sin^2\theta},
    &&\tilde\kappa \approx \frac{\sec\theta\eta\kappa}{\sqrt{\eta^2-\sin^2\theta}}.
    \label{eq253}
    \end{align}
We next introduce the gain coefficient $g$ and its threshold value $g_0$ at resonance frequency
   \begin{align}
   &g:=-2k\kappa = -\frac{4\pi\kappa}{\lambda}, &&g_0:=-2k_0\kappa_0 = -\frac{4\pi\kappa_0}{\lambda_0},\label{gaincoef}
   \end{align}
where $\lambda_0$ corresponds to the resonance wavelength. In the light of (\ref{eq251}), (\ref{eq252}),
(\ref{eq253}) and (\ref{gaincoef}), the spectral singularity condition (\ref{spec-sing}) yields the following
set of equations in the leading order of $\kappa$
  \begin{align}
  &2\tilde{\eta}\,\textrm{Im}\left[\sigma_g\right] \sin\left(2\fK\tilde{\eta}\right) -2\tilde{\eta}^2\cos\left(2\fK\tilde{\eta}\right) = \left[e^{\tilde{g}L}-\left(\tilde{\eta}-\textrm{Re}\left[\sigma_g\right]\right)e^{-\tilde{g}L}\right]\tilde{\kappa}\,\textrm{Im}\left[\sigma_g\right],&\label{spect1}\\
  &2\left[\tilde{\eta}^2+1-\textrm{Re}\left[\sigma_g\right]^2-\textrm{Im}\left[\sigma_g\right]^2\right]\sin\left(2\fK\tilde{\eta}\right) +4\left[\tilde{\eta}\,\textrm{Im}\left[\sigma_g\right]-\tilde{\kappa}\,\textrm{Re}\left[\sigma_g\right]\right]\cos\left(2\fK\tilde{\eta}\right)=\frac{\tilde{\kappa}}{\tilde{\eta}}\left(\fa_{-}e^{-\tilde{g}L}-\fa_{+}e^{\tilde{g}L}\right),\label{spect2}
  \end{align}
where we define $\fa_{\pm}$ and $\tilde{g}$ for convenience as follows
 \begin{align}
 &\fa_{\pm} := \tilde{\eta}^2-1 \pm 2\tilde{\eta}\,\textrm{Re}\left[\sigma_g\right] + \textrm{Re}\left[\sigma_g\right]^2 +\textrm{Im}\left[\sigma_g\right]^2,\nn\\
 &\tilde{g} := \frac{\eta g}{\sqrt{\eta^2 - \sin^2\theta}}. \notag
 \end{align}

In practice, Eqs.~(\ref{spect1}) and (\ref{spect2}) control the lasing behavior of our system in such a way
that the most appropriate system parameters should be accounted for the emergence of optimal impacts. Graphene
effect on the lasing occurrence is revealed by the presence of real and imaginary parts of $\sigma_g$. Thus, a
comprehensive analysis of the involvement of system parameters is required to observe final outcome in the
presence of graphene sheets. For this purpose we exhibit general behaviors of system parameters through the
gain coefficient $g$ plots. We employ Nd:YAG crystals in $\cP\cT$-symmetric bilayer slab system with following
specifications
    \begin{align}
    &\eta = 1.8217, &&\lambda=808~\textrm{nm}, &&L=1~\textrm{cm}, &&\theta = 30^{\circ},\label{specifics}
    \end{align}
and graphene sheets with characterizations
    \begin{align}
    &T = 300~^{\circ}K, &&\Gamma = 0.1~\textrm{meV}, &&\mu = 0.05~\textrm{eV}.\label{graphenespecifics}
    \end{align}

   \begin{figure}[!ht]
   \begin{center}
   \includegraphics[scale=.8]{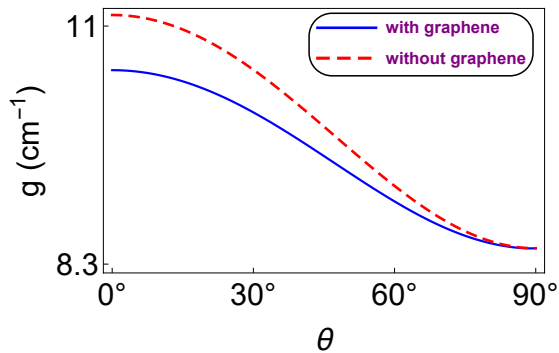}
   \caption{(Color online) Behavior of gain amount $g$ as a function of incidence angle $\theta$ in the cases
of with and without graphene. We use parameter values given in (\ref{specifics}) and (\ref{graphenespecifics})
for our system. Gain amount runs short in the presence of graphene and retains a finite value in the limit of
right angle incidence.}
   \label{fig1m}
   \end{center}
   \end{figure}

In Fig.~\ref{fig1m}, appearance of graphene sheets is distinguished in the plot of gain coefficient $g$ as a
function of incidence angle $\theta$. We use the parameters as given in (\ref{specifics}) for the slab, and in
(\ref{graphenespecifics}) for the graphene component. It is obvious that graphene triggers the gain value to
reduce considerably, depending upon the properties of graphene sheets. As the incidence angle approaches to
$\theta = 90^{\circ}$, graphene effect gets trivial and amount of the gain coefficient survives at a finite
value as distinct from an individual layer case alone.

\begin{figure}[!h]
    \begin{center}
    \includegraphics[scale=.8]{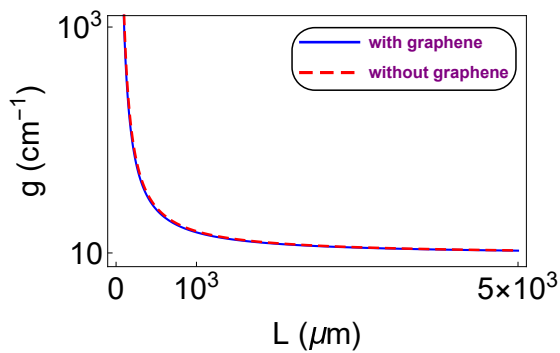}
    \caption{(Color online) Gain value $g$ as a function of slab thickness $L$ when parameters of the system
are given by (\ref{specifics}) and (\ref{graphenespecifics}). The required lasing gain amount slightly
descends for smaller slabs, especially at micron sizes.}
    \label{fig2m}
    \end{center}
    \end{figure}

Figure~\ref{fig2m} reflects essential behavior of curves in the plane of slab thickness $L$ and gain $g$ once
the graphene sheets are inserted into the system. Gain values slightly drop off especially more at smaller
thicknesses of micron sizes.

\begin{figure}[!h]
    \begin{center}
    \includegraphics[scale=.75]{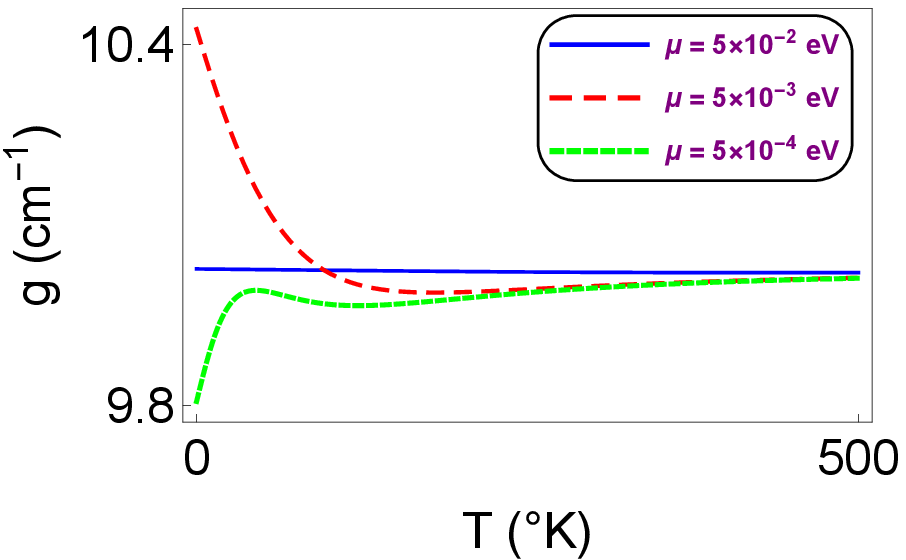} \\
    \includegraphics[scale=.75]{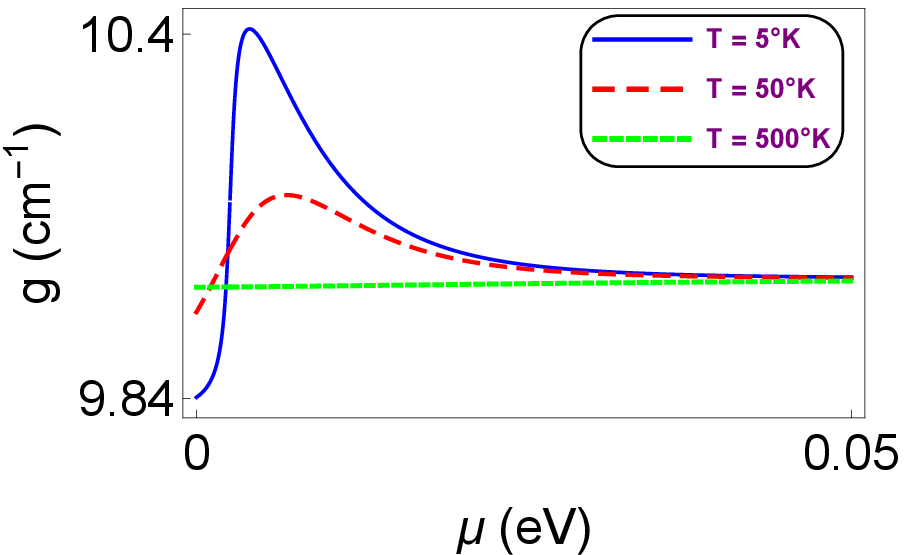}
    \caption{(Color online) The roles of temperature and chemical potential of graphene sheets on the gain
reduction. Temperatures around the absolute zero together with very small chemical potentials yield a large
decrement in the gain coefficient.}
    \label{fig3m}
    \end{center}
    \end{figure}

It turns out that characteristics of graphene layers manifest themselves in determining the rate of gain
decrement. Fig.~\ref{fig3m} displays attitudes of temperature and chemical potential of graphene sheets. It
is revealed that optimal reduction of gain amount is achieved by employing as much lower temperatures and
chemical potentials as possible. In particular, chemical potentials less than about $\mu = 3$~meV are
favorable for the best effect. This is the case $\mu \leq \hbar \omega/2$. Around the resonance of graphene
where the relation $\mu = \hbar \omega/2$ holds, the required gain amount for lasing rises and then proceeds
to decay to $g \approx 10.02~\textrm{cm}^{-1}$ when $\mu \geq \hbar \omega/2$. Lastly, we remark that
temperature (also chemical potential) dependence of the refractive indices is ignored safely since it yields
a negligible effect (about 0.001\%) within the temperature limits of interest~\cite{temp,temp2}, which
guarantees validity of our results.

\section{Presence of Dispersion Content}

\begin{figure}[!h]
    \begin{center}
    \includegraphics[scale=.5]{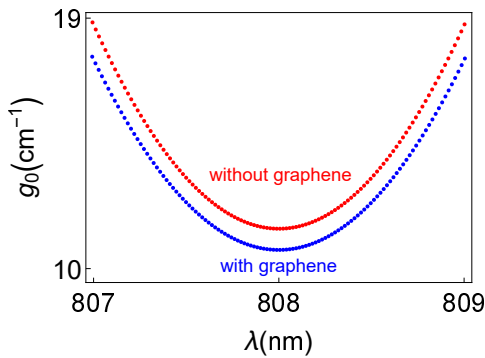}
    \caption{(Color online) Threshold gain $g_0$ versus wavelength $\lambda$ behavior corresponding to with
and without graphene cases in the presence of dispersion. Apparently, $g_0$ has smaller values when the
graphene sheets are inserted.}
    \label{figdispersion}
    \end{center}
    \end{figure}

We now assume that there exist a dispersion in the refractive index $\fn$, and investigate its possible
impacts on the spectral singularities. For this purpose, we include wavenumber $k$ dependence of $\fn$. We
assume that active part of our optical system composing the gain ingredient is formed by doping a host
medium of refractive index $n_0$, and its refractive index satisfies the following dispersion relation
    \be
    \fn^2= n_0^2-
    \frac{\hat\omega_p^2}{\hat\omega^2-1+i\hat\gamma\,\hat\omega}.
    \label{epsilon}
    \ee
Here $\hat\omega:=\omega/\omega_0$, $\hat\gamma:=\gamma/\omega_0$, $\hat\omega_p:=\omega_p/\omega_0$,
$\omega_0$ is the resonance frequency, $\gamma$ is the damping coefficient, and $\omega_p$ is the plasmon
frequency. The $\hat\omega_p^2$ can be described in leading order of the imaginary part $\kappa_0$ of $\fn$
at the resonance wavelength $\lambda_0:=2\pi c/\omega_0$, by the expression
$\hat\omega_p^2=2n_0\hat\gamma\kappa_0$~\cite{pra-2011a}. After inserting this relation into (\ref{epsilon}),
employing the first expression of (\ref{eq252}), and ignoring quadratic and higher order terms in $\kappa_0$,
we obtain the real and imaginary parts of the refractive index as~\cite{CPA,CPA-1,CPA-2,CPA-3}
     \begin{align}
    &\eta\approx n_0+\frac{\kappa_0\hat\gamma(1-\hat\omega^2)}{(1-\hat\omega^2)^2+
    \hat\gamma^2\hat\omega^2},
    &&\kappa\approx\frac{\kappa_0\hat\gamma^2\hat\omega}{(1-\hat\omega^2)^2+
    \hat\gamma^2\hat\omega^2}.
    \label{eqz301}
    \end{align}
At resonance wavelength $\lambda_0$, the $\kappa_0$ can be written as $\kappa_0=-\lambda_{0}g_0/4\pi$,
see~(\ref{gaincoef}). Substituting this relation in (\ref{eqz301}) and making use of (\ref{eq252}) and
(\ref{spec-sing}), we can determine $\lambda$ and $g_{0}$ values for the spectral singularities. These are
explicitly shown in the $\lambda$-$g_0$ plane in Fig.~\ref{figdispersion} for our setup with slab and
graphene properties listed in (\ref{specifics}) and (\ref{graphenespecifics}), and for the incidence angle
$\theta=0^{\circ}$. Furthermore, Nd:YAG crystals which form the slab material hold the following
$\hat\gamma$ value for given $n_0$ and $\lambda_0$ values~\cite{silfvast}
    \begin{align}
    &n_0=1.8217, &&\lambda_0=808\,{\rm nm}, && \hat\gamma=0.003094.
    \label{specifications}
    \end{align}

It appears that graphene sheets with associated parameters lead to shift down in the location of spectral
singularity points. This verifies our findings explored in the previous section. Again at temperatures close
to absolute zero and chemical potentials much lower values, the spectral singularity points move faster down
in the $\lambda$-$g_0$ plane.

\section{CPA Laser Action}

\begin{figure*}[!ht]
    \begin{center}
    \includegraphics[scale=.48]{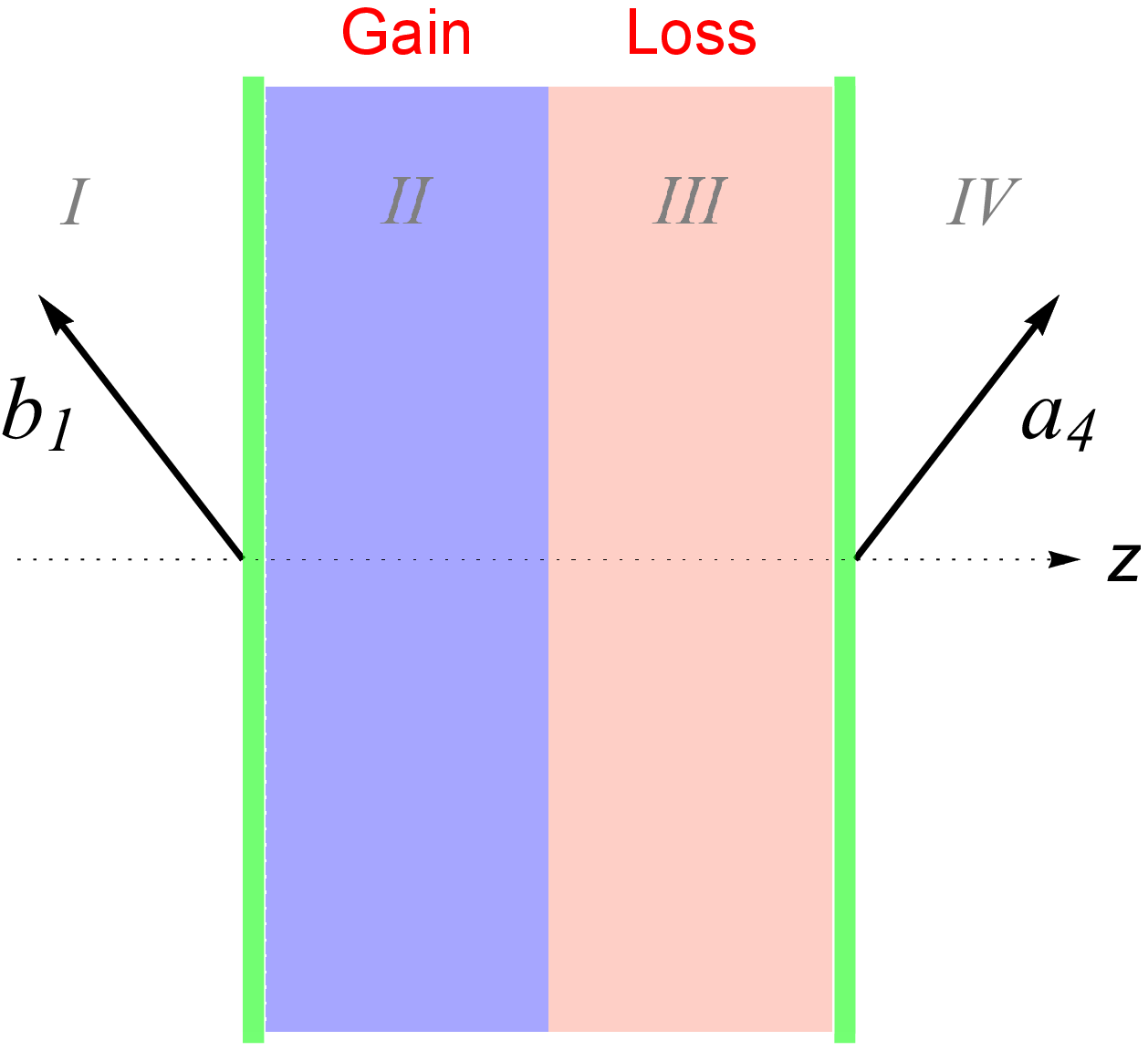} ~~~~~~~~~~~~~
    \includegraphics[scale=.48]{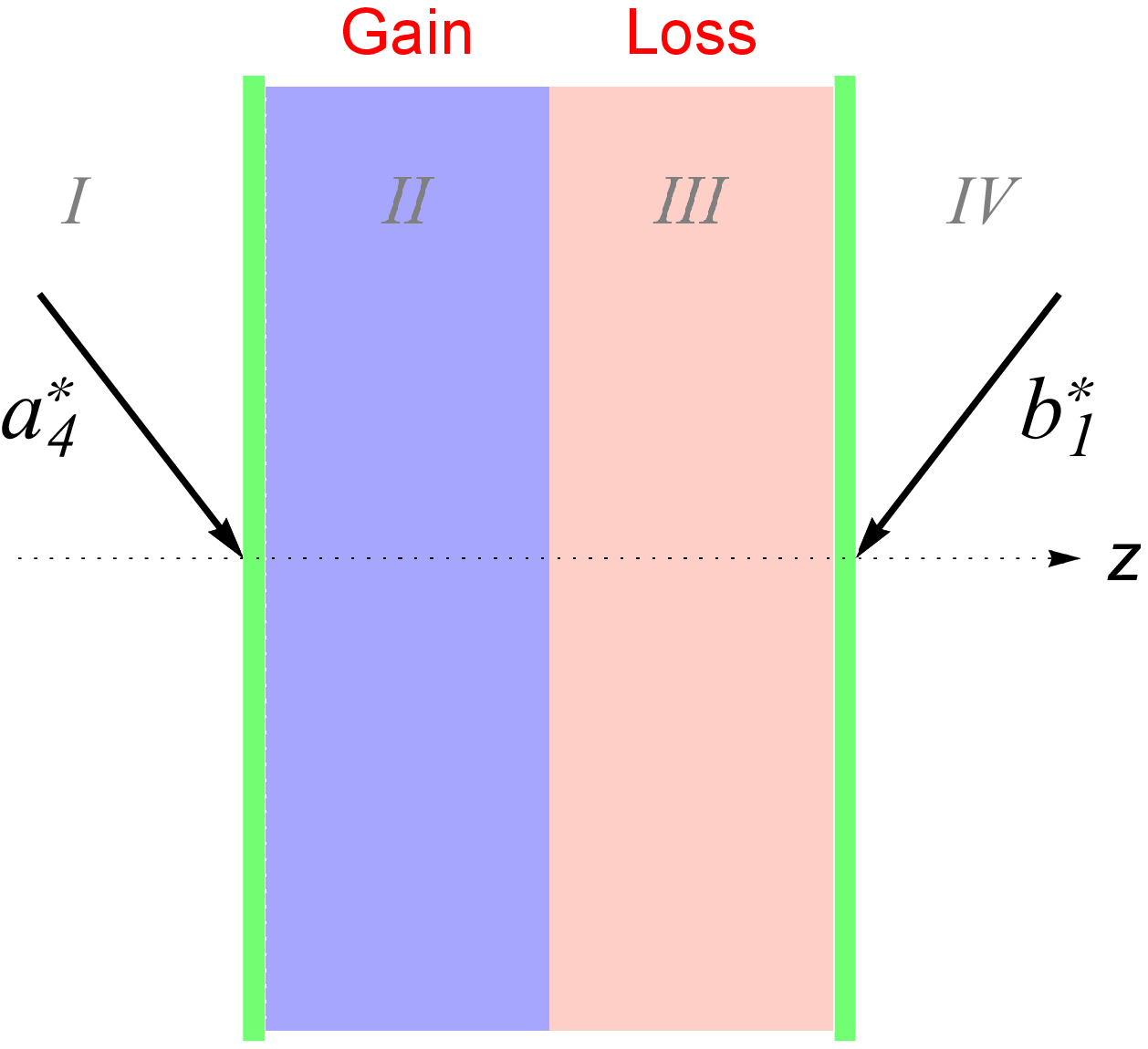}
    \caption{(Color online) Configurations representing the spectral singularities (left plot), and coherent
perfect absorber (CPA) (right plot). Spectral singularities correspond to purely outgoing waves while a CPA
forms only when correct phases and amplitudes of incoming waves, as defined in Eqs.~\ref{rho2=} and
\ref{phi2=}, are generated.}
    \label{figCPA}
    \end{center}
    \end{figure*}

Our $\cP\cT$-symmetric optical slab encrusted by graphene sheets serves as a CPA provided that time reversed
system is fulfilled once the spectral singularities are prevalent. This phenomenon happens to exist only if
correct phase and amplitude of incoming waves are originated. Thus, incoming waves are perfectly absorbed by
the optical system that gives rise to a CPA-laser, see Fig.~\ref{figCPA} for a pictorial demonstration of
the phenomenon.

Spectral singularities displayed in the left panel of Fig.~\ref{figCPA} describe purely outgoing waves. Time
reversed case is just obtained by complex conjugation of waves outside the slab such that gain and loss parts
are interchanged. To make a CPA-laser, we switch the gain and loss parts of time reversed case so that desired
incoming waves can entirely be obtained as shown in the right panel of Fig.~\ref{figCPA}. For a required
phenomena, waves outside the active part of the optical slab can be expressed as in Table~\ref{table02} in
terms of solutions of Maxwell equations in the exterior regions as given in (\ref{ez1}) and (\ref{E-theta}).

\begin{table*}[!h]
\centering
{%
\begin{tabular}{@{\extracolsep{4pt}}llcccccc}
\toprule
{} & {} & \multicolumn{1}{c}{Spectrally Singular Waves} & \multicolumn{1}{c}{Time Reversed Waves} &\multicolumn{1}{c}{CPA Waves}\\
 \cline{2-2}
 \cline{3-5}
 \cline{6-8}
   \hline
  & Region $I$    & $b_1\,e^{i\fK (\mathbf{x} \tan\theta - \mathbf{z})}$ & $b_1^{\ast}\,e^{-i\fK (\mathbf{x} \tan\theta - \mathbf{z})}$ & $a_4^{\ast}\,e^{-2i\fK}\,e^{-i\fK (\mathbf{x} \tan\theta + \mathbf{z})}$ \\
  & Region $I\!V$ & $a_4\,e^{i\fK (\mathbf{x} \tan\theta + \mathbf{z})}$ & $a_4^{\ast}\,e^{-i\fK (\mathbf{x} \tan\theta + \mathbf{z})}$ & $b_1^{\ast}\,e^{2i\fK}\,e^{-i\fK (\mathbf{x} \tan\theta - \mathbf{z})}$ \\
 \hline
\end{tabular}%
}
\caption{Waves outside the $\cP\cT$-symmetric optical slab with graphene corresponding to phenomena specified
in the first row.} \label{table02}
\end{table*}

\begin{figure}[!ht]
    \begin{center}
    \includegraphics[scale=.46]{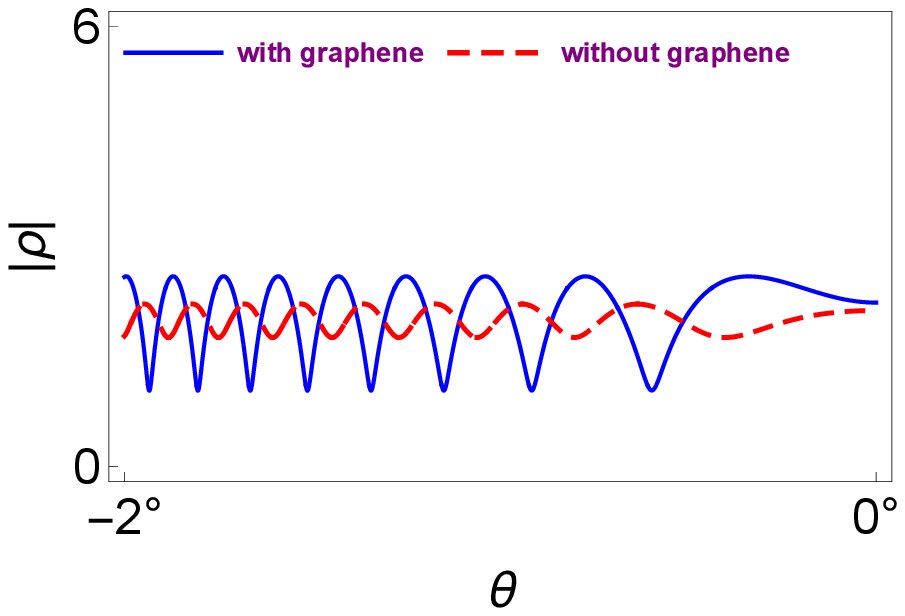}
    \includegraphics[scale=.5]{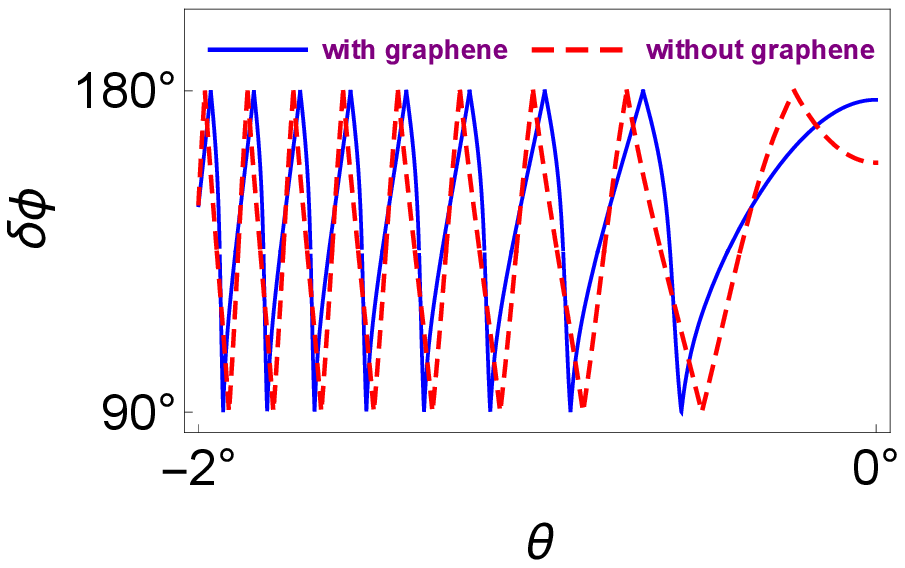}\\
    \includegraphics[scale=.46]{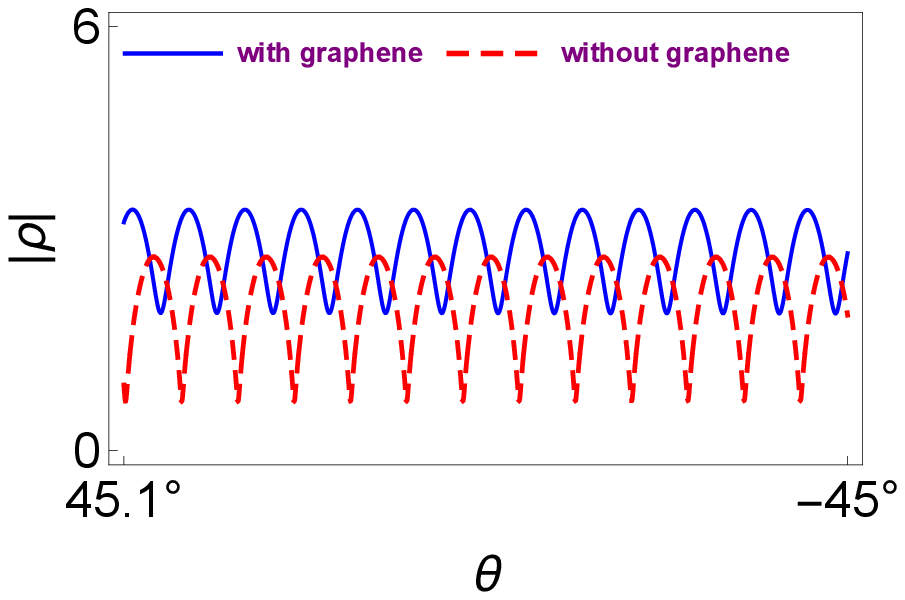}
    \includegraphics[scale=.5]{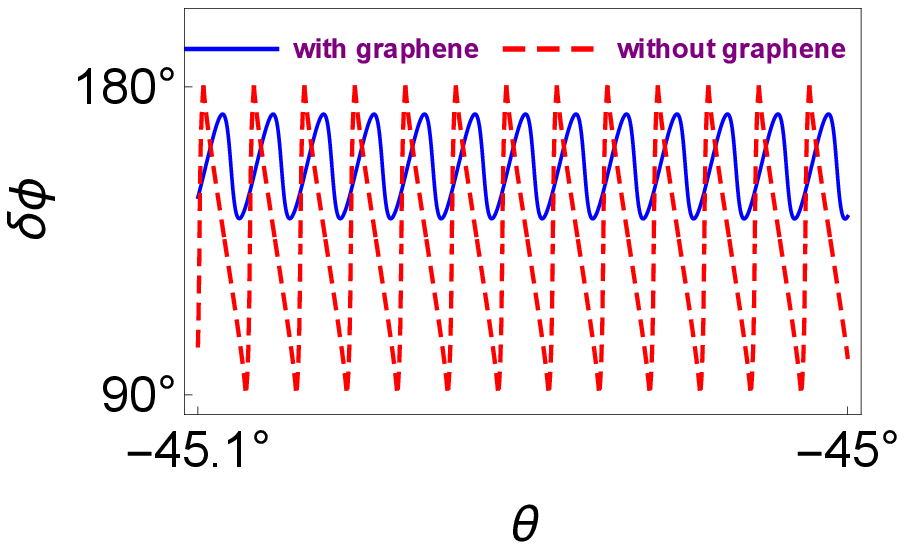}\\
    \includegraphics[scale=.46]{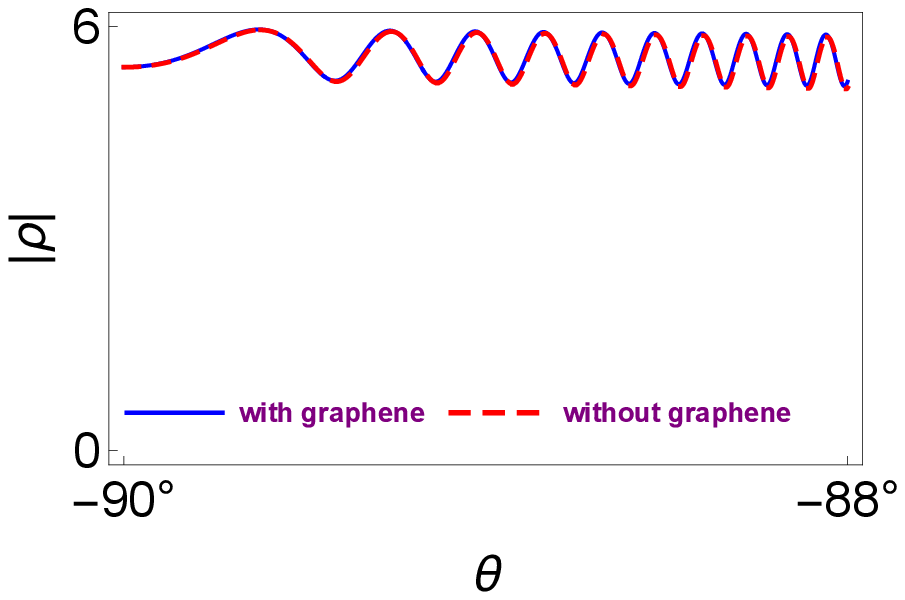}
    \includegraphics[scale=.5]{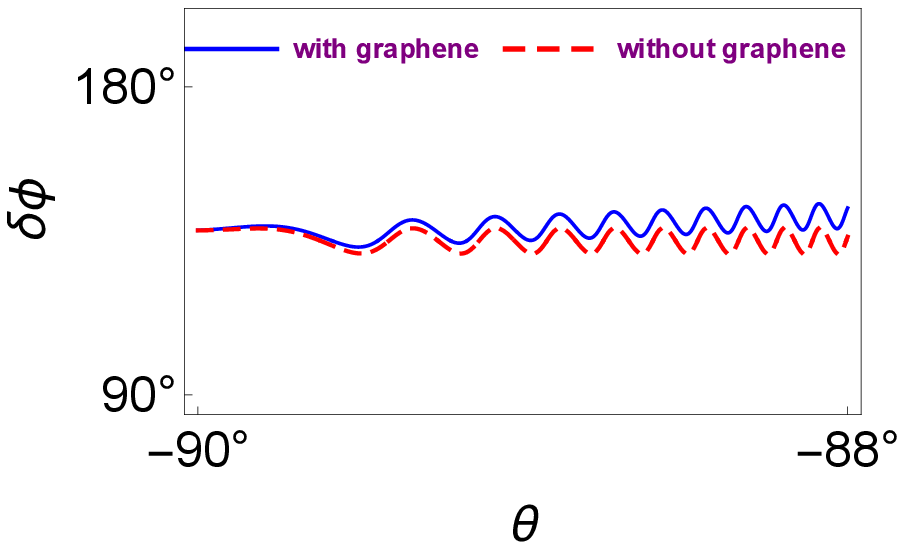}\\
    \includegraphics[scale=.45]{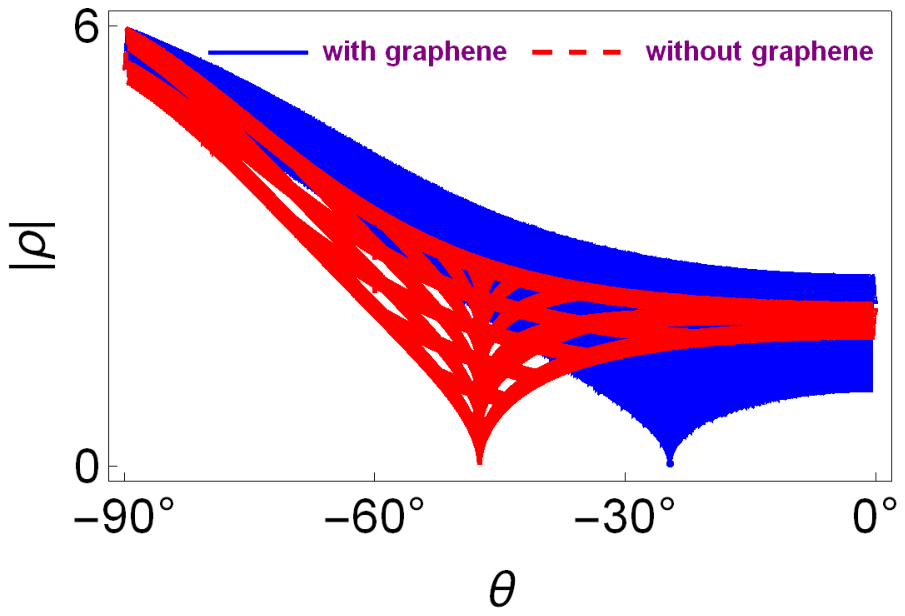}
    \includegraphics[scale=.5]{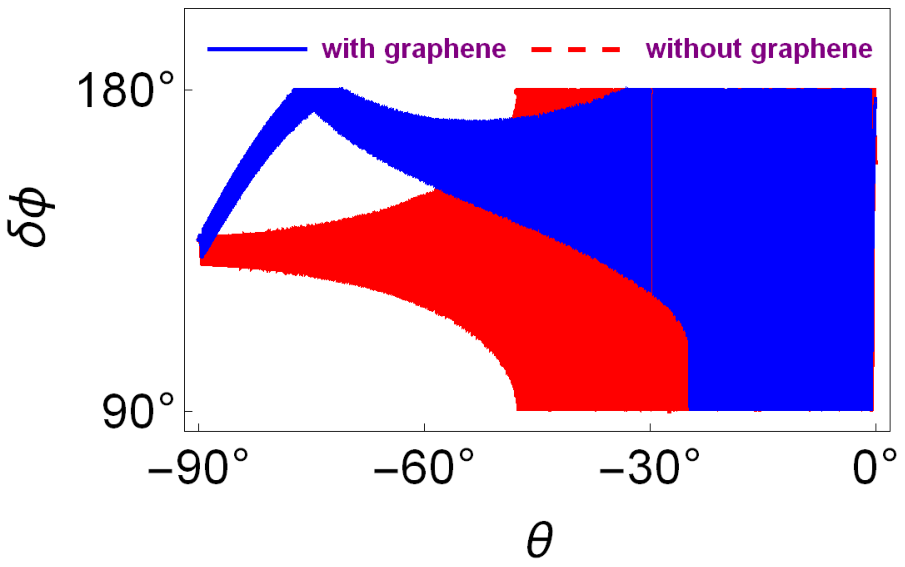}
    \caption{(Color online) Dependence of $\left|\rho\right|$ and $\delta\phi$ on incidence angle $\theta$
in a system with and without graphene. System parameters are taken as in (\ref{specifics}) and
(\ref{graphenespecifics}). Notice that presence of graphene results in larger $\left|\rho\right|$ and
$\delta\phi$ depending on the $\theta$.}
    \label{rho=phi=theta}
    \end{center}
    \end{figure}

\begin{figure}[!ht]
    \begin{center}
    \includegraphics[scale=.77]{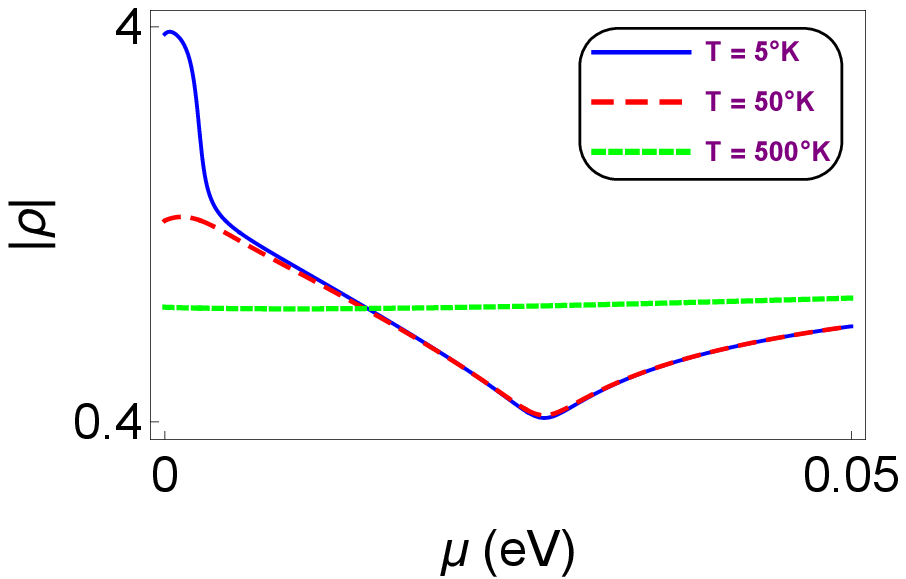}
    \includegraphics[scale=.77]{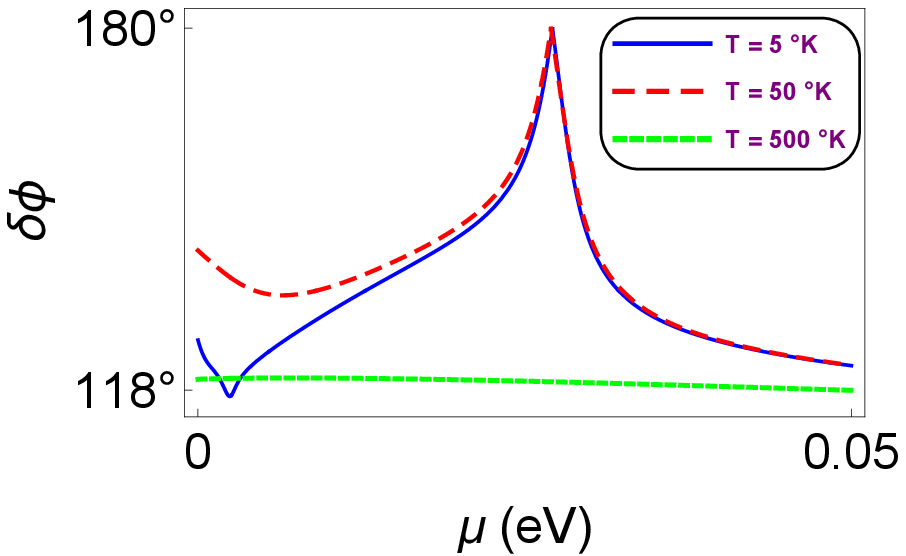}\\
    \caption{(Color online) Conditions for obtaining a CPA-laser on how to tune the chemical potential of
graphene at various temperatures. Incidence angle is opted for $\theta=-30^{\circ}$. Graphene manifests
itself at very small $\mu$ values. Increasing temperature leads $\left|\rho\right|$ to move down, and
$\delta\phi$ to move up at constant $\mu$.}
    \label{rho=phi=temp1}
    \end{center}
    \end{figure}

\begin{figure}[!hb]
    \begin{center}
    \includegraphics[scale=.77]{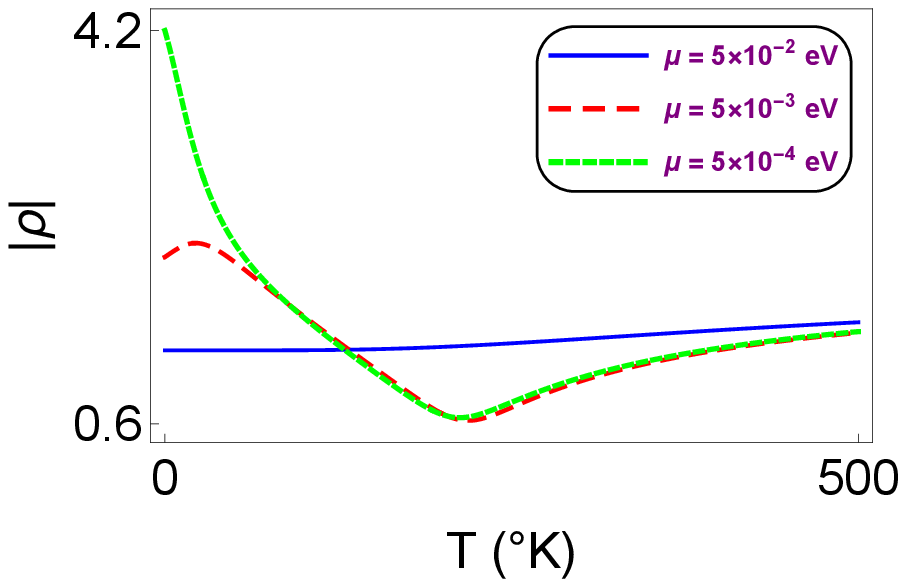}
    \includegraphics[scale=.77]{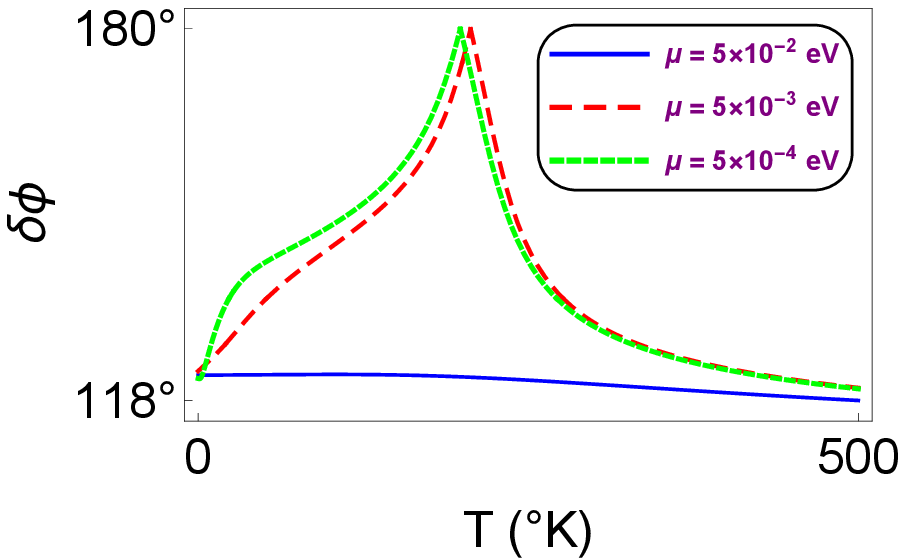}\\
    \caption{(Color online) Impact of graphene temperature on $\left|\rho\right|$ and $\delta\phi$ necessary
for obtaining a CPA-laser. Incidence angle is set to $\theta = -30^{\circ}$. Temperatures close to resonance
values $T \approx 220~^{\circ}\textrm{K}$ together with small $\mu$ values are favorable if one desires to
feel the presence of graphene in a CPA-laser.}
    \label{rho=phi=temp2}
    \end{center}
    \end{figure}

CPA-laser operates once the incoming waves emergent by angle $-\theta$ are absorbed perfectly so that full
destructive interference occurs. This can be measured by the ratio $\rho$ of complex amplitude of incoming
waves for $z \rightarrow 0$ and $z \rightarrow 2L$. In view of amplitudes given in Table~\ref{table02}, this
is expressed as
\be
\rho = \frac{a_4^{\ast}\,e^{-2i\fK}}{b_1^{\ast}}. \label{rho=}
\ee
In fact, $a_4$ can be denominated in terms of $b_1$ by employing the spectral singularity condition. We
recall that spectral singularities correspond to purely outgoing waves such that
\be
a_1 = b_4 = 0. \notag
\ee
This, together with the boundary conditions in Table~\ref{table01} and spectral singularity condition in
(\ref{spec-sing}), leads to
\be
a_4 =\frac{e^{-2i\fK}\,b_1}{\tilde{\fn}^{\ast}} \sqrt{\frac{V_{+}\,V_{-}^{\ast}}
{\left[1 + \fu_{+}^{\ast}\right]\left[1 - \fu_{+}\right]}}. \label{a4=}
\ee
Thus, it is easy to show that incoming waves are perfectly absorbed provided that ratio of incoming
amplitudes specified by $\rho$ satisfies
\be
\rho = \frac{1}{\tilde{\fn}}\sqrt{\frac{V_{+}^{\ast}\,V_{-}}{\left[1 + \fu_{+}\right]
\left[1 - \fu_{+}^{\ast}\right]}}. \label{rho2=}
\ee
Hence, ratio of amplitudes and phase factors of the waves incoming from left- and right-hand sides are
characterized by $\left|\rho\right|$ and $\delta\phi$ respectively, and the latter is quantified as
\be
e^{i\delta\phi} = \frac{\rho}{\left|\rho\right|}.\label{phi2=}
\ee
Consequently, the CPA-laser action can outrightly be obtained once the $\left|\rho\right|$ and $\delta\phi$
of incoming waves from the left and right-hand sides are tuned in according to expressions (\ref{rho2=}) and
(\ref{phi2=}) respectively. Pictorial representation of how these quantities are influenced by parameters
of the optical system are clearly shown in Figs.~\ref{rho=phi=theta}, \ref{rho=phi=temp1} and
\ref{rho=phi=temp2}. Note that in these figures the parameters assume the values given in (\ref{specifics})
and (\ref{graphenespecifics}).

Figure~\ref{rho=phi=theta} displays dependence of $\left|\rho\right|$ and $\delta\phi$ on incidence angle
$\theta$. We set the gain amount to $g=10~\textrm{cm}^{-1}$, $\lambda=808$~nm, and $\mu=5$~meV. We notice
that $\left|\rho\right|$ and $\delta\phi$ oscillate with angle $\theta$ such that peaks in the presence of
graphene case slightly shift with respect to without graphene case, especially for small incidence angles,
and they almost coincide for large incidence angles. It is obvious that presence of graphene requires larger
$\left|\rho\right|$ compared to without graphene case for small and moderate incidence angles, and this
influence decreases as $\theta$ gets larger. This means that amplitude of wave coming from the left side
should be adjusted to have a little higher value compared to the amplitude of waves coming from the right
side. In a similar manner, phase difference factor $\delta\phi$ for graphene shifts slightly, and could get
any value for small incidence angles, and graphene induces this phase difference to rise significantly as
the incidence angle increases. This amounts to that graphene mostly yields almost destructive interference.
Graphene effect is not felt much for the angles close to right angles. See the last row of
Fig.~\ref{rho=phi=theta} for overall view of incidence angle dependence.

In Figs.~\ref{rho=phi=temp1} and \ref{rho=phi=temp2}, one explicitly realizes the effect of graphene for
obtaining a CPA-laser. We employ the incidence angle $\theta=-30^{\circ}$, gain value $g=10~\textrm{cm}^{-1}$,
and wavelength $\lambda=808$~nm. Figure~\ref{rho=phi=temp1} demonstrates how chemical potential variation of
graphene sheets impacts on the ratio of amplitudes $\left|\rho\right|$ and phase difference $\delta\phi$
corresponding to different temperatures. We observe that chemical potentials less than about $\mu=5$~meV give
more $\left|\rho\right|$ and relatively small phase difference, especially at smaller temperatures. When
$\mu$ increases till about 0.028~eV, the  $\left|\rho\right|$ decreases and $\delta\phi$ increases. The point
$\mu \approx 0.028$~eV corresponds to resonance value of graphene so that it is felt at small temperatures as
noticed in the figure. When chemical potential gets values higher than $\mu = 0.028$~eV, $\left|\rho\right|$
starts to scale up to get asymptotic value of about $\left|\rho\right|\approx 1.4$ while $\delta\phi$ reduces
to get a value around $\delta\phi\approx 118^{\circ}$. These findings are verified in Fig.~\ref{rho=phi=temp2},
i.e. the effect of graphene is perceived at considerably small temperatures, and temperatures giving rise to
the resonance effect particularly for small chemical potentials such that $\left|\rho\right|$ increases
whereas $\delta\phi$ decreases as temperature decreases from $T=50~^{\circ}\textrm{K}$, and
$\left|\rho\right|$ decreases to minimum value and $\delta\phi$ increases to maximum value in temperature
range $50~^{\circ}\textrm{K} \lesssim T \lesssim 220~^{\circ}\textrm{K}$. It is understood that temperature
and chemical potential values close to resonance effect of graphene are favorable in order to get a
well-adjusted CPA-laser.

Table~\ref{table3} demonstrates some parameter values belonging to our optical system with and without
graphene to realize a CPA-laser as an instructive guide for the experimental attempts. Properties of graphene
sheets are specified by temperature $ T=300~^{\circ}\textrm{K}$, chemical potential of $\mu=0.05$~eV and
$\Gamma=0.1$~meV. Table~\ref{table4} explicitly presents values necessary to build a graphene based CPA-laser
corresponding to graphene parameter values $T=50~^{\circ}\textrm{K}$ and $\mu=0.027$~eV (left board), and
$T=200~^{\circ}\textrm{K}$ and $\mu=0.5$~eV (right board).

 \begin{table*}[ht]
\centering
{%
\begin{tabular}{@{\extracolsep{4pt}}llcccccccc}
\toprule
{} & {} & {} & \multicolumn{3}{c}{with Graphene}  & {$-$} &\multicolumn{3}{c}{without Graphene}\\
 \cline{2-2}
 \cline{3-5}
 \cline{6-8}
   \hline
  & $\theta$ & {} & $0^{\circ}$ & $-40^{\circ}$ & $-80^{\circ}$ & {} & $0^{\circ}$ & $-40^{\circ}$ & $-80^{\circ}$ \\
  & $\lambda$ & {} & $807.996~\textrm{nm}$ & $808.008~\textrm{nm}$ & $807.993~\textrm{nm}$ & {} & $808.006~\textrm{nm}$ & $808.001~\textrm{nm}$ & $808.009~\textrm{nm}$ \\
  & $g$ & {} & $10.859~\textrm{cm}^{-1}$ & $10.104~\textrm{cm}^{-1}$ & $9.002~\textrm{cm}^{-1}$ & {} & $11.433~\textrm{cm}^{-1}$ & $10.458~\textrm{cm}^{-1}$ & $9.025~\textrm{cm}^{-1}$ \\
  & $\kappa$ & {} & $-6.982\times 10^{-5}$ & $-6.496\times 10^{-5}$ & $-5.788\times 10^{-5}$ & {} & $-7.351\times 10^{-5}$ & $-6.724\times 10^{-5}$ & $-5.803\times 10^{-5}$ \\
  & $\left|\rho\right|$ & {} & 1.9053 & 1.8883 & 1.7817 & {} & $0.5824$ & $0.7597$ & $1.5929$ \\
  & $\delta\phi$ & {} & $146.106^{\circ}$ & $148.620^{\circ}$ & $179.940^{\circ}$ & {} & $135.008^{\circ}$ & $135.007^{\circ}$ & $135.004^{\circ}$ \\
 \hline
\end{tabular}%
}
\caption{Physical parameters of a CPA construct for various incidence angles corresponding to with and
without graphene cases. Here graphene has $T=300~^{\circ}\textrm{K}$, $\mu=0.05$~eV, and $\Gamma=0.1$~meV.}
\label{table3}
\end{table*}

 \begin{table*}[ht]
\centering
{%
\begin{tabular}{@{\extracolsep{4pt}}llcccccccc}
\toprule
{} & {} & {} & \multicolumn{3}{c}{$T = 50~^{\circ}$ and $\mu = 0.027~eV$}  & {$-$} &\multicolumn{3}{c}{$T = 200~^{\circ}$ and $\mu = 0.0005~eV$}\\
 \cline{2-2}
 \cline{3-5}
 \cline{6-8}
   \hline
  & $\theta$ & {} & $0^{\circ}$ & $-40^{\circ}$ & $-80^{\circ}$ & {} & $0^{\circ}$ & $-40^{\circ}$ & $-80^{\circ}$ \\
  & $\lambda$ & {} & $807.995~\textrm{nm}$ & $808.007~\textrm{nm}$ & $807.991~\textrm{nm}$ & {} & $807.994~\textrm{nm}$ & $808.007~\textrm{nm}$ & $807.991~\textrm{nm}$ \\
  & $g$ & {} & $10.872~\textrm{cm}^{-1}$ & $10.119~\textrm{cm}^{-1}$ & $9.014~\textrm{cm}^{-1}$ & {} & $10.819~\textrm{cm}^{-1}$ & $10.064~\textrm{cm}^{-1}$ & $8.971~\textrm{cm}^{-1}$ \\
  & $\kappa$ & {} & $-6.990\times 10^{-5}$ & $-6.506\times 10^{-5}$ & $-5.796\times 10^{-5}$ & {} & $-6.956\times 10^{-5}$ & $-6.471\times 10^{-5}$ & $-5.768\times 10^{-5}$ \\
  & $\left|\rho\right|$ & {} & 1.6713 & 1.6324 & 1.6591 & {} & $1.6291$ & $1.5982$ & $1.6632$ \\
  & $\delta\phi$ & {} & $157.613^{\circ}$ & $162.139^{\circ}$ & $160.031^{\circ}$ & {} & $153.293^{\circ}$ & $158.044^{\circ}$ & $160.507^{\circ}$ \\
 \hline
\end{tabular}%
}
\caption{Physical parameters of a graphene-based CPA for various incidence angles. Here, graphene parameters
are $T=50~^{\circ}\textrm{K}$, $\mu=0.027$~eV, and $T=200~^{\circ}\textrm{K}$, $\mu = 0.5$~meV respectively.
Scattering rate is taken as $\Gamma = 0.1$~meV.} \label{table4}
\end{table*}

\section{Concluding Remarks}\label{S9}

This study benefits the idea that a CPA-laser is the time reversal construct of a regular laser which could
be expressed by means of the spectral singularities. Nowadays, experimental realization of CPA-lasers is the
main challenge due to difficulties in adjusting exact amplitude and phase factors of incoming waves. Current
studies exploit various techniques in order to measure the quantities pointing CPA actions. In this work, we
employed an optical system which respects the property of $\cP\cT$-symmetry accompanied by graphene
containment. In \cite{lastpaper}, necessary and sufficient conditions for implementing a CPA-laser based on
the feature of $\cP\cT$-symmetry are given, and in this current study we performed a comprehensive analysis
aiming to find the conditions for CPA-laser action based on graphene sheets. Hence, the results of this
study guide experimental attempts for realization of $\cP\cT$-symmetric CPA phenomenon with graphene.

We made use of distinctive traits of the transfer matrix formalism and determined the spectral singularities,
which give rise to lasing threshold condition reflecting the presence of graphene in a $\cP\cT$-symmetric
optical system. The transfer matrix approach grounds the power of boundary conditions arising from the
solutions coming directly from Maxwell equations. Table~\ref{table01} reveals the presence of graphene
sheets in boundary conditions in the form of complex function $\fu_{\pm}^{(j)}$, see Eq.~(\ref{u=}). We
derived exact expressions for the conditions of lasing threshold in (\ref{spec-sing}) and of coherent
perfect absorber in (\ref{rho2=}) and (\ref{phi2=}). We employ a perturbative approach to obtain optimal
conditions arising from the system parameters.

We require that graphene coatings respect overall $\cP\cT$ symmetry which leads to formation of currents
flowing in opposite directions. We emphasize that $\cP\cT$ symmetry enables the emergence of lasing and
CPA conditions due to its power to control the system parameters as distinct from non-$\cP\cT$-symmetric
structures. In particular, we find out that graphene insertion into a $\cP\cT$-symmetric optical system
leads lower gain values depending upon the graphene features, especially temperature and chemical potential.
At this point, resonance occurrence of graphene plays a significant role which happens to exist at
$\mu_r \approx 2.56$~meV at absolute zero temperature, and slightly shifts up with increasing temperature.
In general, below the resonance point, the lower temperature and chemical potential of graphene sheets are,
the less gain amount is. The results of spectral singularities and in turn lasing threshold conditions
facilitate the path towards obtaining a CPA because of coincidence of both phenomena at the same points.

We observe that efficiency of a CPA-laser could be improved once the parameters of the optical system
together with graphene features are well-adjusted. We consider $\cP\cT$-symmetry to achieve computational
and thus experimental accessibility of the optical system by tuning appropriate parameters of the system.
Accordingly, we place graphene sheets at the ends of $\cP\cT$-symmetric slab system just to realize that
optimal conditions for the CPA-laser action can be obtained by computing appropriate intensity and phase
contrasts. In order to get a lower intensity and higher phase contrast that yield an almost destructive
interference, one should opt for lower temperatures and chemical potentials for the graphene sheets at
which the resonance effect occurs. Also, small incidence angles provide relatively lower intensities in the
presence of graphene, which lead to larger phase difference as desired. This is the main reason to use
graphene sheets in the optical system.

In view of our findings, one can shape a reliable CPA equipment provided that parameters specifying the
graphene and bilayer slab system are well-adjusted. In Tables~\ref{table3} and \ref{table4} we provide
explicit values of parameters in order to build a concrete CPA. Our primary purpose in placing graphene is
to improve absorption of waves and utilize arrangement of parameters. We explored that this is achieved at
values especially around the resonance effect of graphene. Furthermore, it is observed that graphene causes
the necessary gain amount to lessen. In this respect, we infer that presence of graphene gives valuable
information about enhancement of absorption in a CPA, which helps building a better CPA. This suggests even
a more effective material could be used instead of graphene to make a unity ratio of amplitudes and smooth
phase contrast to make a perfect destructive interference. In this direction recently some prominent
candidates have been intensely studied like Weyl semimetals may be in the limelight to build a better CPA.

As a final note, it would be interesting to describe the role of surface plasmon polaritons (SPPs) in our system. But, no surface modes should exist at the interface between the slab and graphene since we focused on TE polarization solutions. SPPs exist only for TM
polarization.~\cite{Maier}.\\[6pt]

\end{document}